\documentclass[12pt]{article}
\usepackage{a4wide,amsmath,amssymb,array}

\catcode`\_=12
\let\_=_
\catcode`\!=8

\newcommand{\ie}{i.e.\ }
\newcommand{\eg}{e.g.\ }
\newcommand{\ran}[3]{\ensuremath{#1 = #2\dots #3}}

\newenvironment{slhablocks}{%
  \begin{center}
  \begin{tt}
  \begin{small}
  \begin{tabular}{|>{\normalfont\scshape}l|l|ll|} \hline
  \textnormal{Block name} &
  \textnormal{Array and length} &
  \textnormal{Members} & \\
  \hline\hline
}{%
  \hline
  \end{tabular}
  \end{small}
  \end{tt}
  \end{center}
}

\makeatletter
\def\reportno#1{\gdef\@reportno{#1}}
\def\@maketitle{%
  \hfill{\small\begin{tabular}[t]{r}%
    \@reportno
  \end{tabular}\par}%
  \vskip 2em%
  \begin{center}%
    \let \footnote \thanks
    {\large \@title \par}%
    \vskip 1.5em%
    {
      \lineskip .5em%
      \begin{tabular}[t]{c}%
        \@author  
      \end{tabular}\par}%
    \vskip 1em%
    {
     \@date}%
  \end{center}%
  \par
  \vskip 1.5em}
\makeatother

\begin{document}

\reportno{MPP--2006--46\\hep--ph/0605049}

\title{SUSY Les Houches Accord 2 I/O made easy}

\author{T. Hahn \\
Max-Planck-Institut f\"ur Physik \\
F\"ohringer Ring 6, D--80805 Munich, Germany}

\date{Oct 18, 2006}

\maketitle

\begin{abstract}
A library for reading and writing data in the SUSY Les Houches Accord 2
format is presented.  The implementation is in native Fortran 77.  The 
data are contained in a single array conveniently indexed by 
preprocessor statements.
\end{abstract}


\section{Introduction}

The original SUSY Les Houches Accord \cite{slha1} (SLHA1 in the
following) has standardized and significantly simplified the exchange of
MSSM input and output parameters between such disparate applications as
spectrum calculators and event generators.  Meanwhile, agreement has
been reached also about the encoding of many extensions of the MSSM
which has led to a preliminary SLHA2 document \cite{slha2}.

While the SLHA specifications include the precise formats for Fortran
I/O, it is nevertheless not entirely straightforward to read or write a
file in SLHA format.  The present library provides the user with simple
routines to read and write files in SLHA format, as well as a few
utility routines.  One thing the library does not do is modify the
numbers, which means there is no routine to compute, say, a particular
quantity at a new scale.  The data structures and subroutines are
set up such that only very few changes are necessary when upgrading from 
the SLHALib 1 \cite{slhalib1}.

Sect.~\ref{sect:data} describes the organization of the data
structures, Sect.~\ref{sect:ref} gives the reference information for
the library routines, Sect.~\ref{sect:examples} shows the usage in some
examples, Sect.~\ref{sect:build} contains download and build
instructions, and Sect.~\ref{sect:summary} summarizes.


\section{Data structures}
\label{sect:data}

The SLHA library is written in Fortran 77.  All routines operate on a
double complex array, \texttt{slhadata}, which is about the simplest
conceivable data format for this purpose in Fortran.  For convenience of
use, this array is accessed via preprocessor statements, so the user
never needs to memorize any actual indices for the \texttt{slhadata}
array.  A file containing the preprocessor definitions must thus be
included.

The \texttt{slhadata} array consists of a `static' part containing the
information from SLHA \texttt{BLOCK} sections and a `dynamic' part
containing the information from SLHA \texttt{DECAY} sections.  The
static part is indexed by preprocessor variables defined in
\texttt{SLHA.h}, the dynamic part is accessed through the
\texttt{SLHANewDecay}, \texttt{SLHAFindDecay}, \texttt{SLHAAddDecay},
\texttt{SLHAGetDecay}, and \texttt{SLHADecayTable} functions and
subroutines (see Sect.~\ref{sect:ref}).

In addition, descriptive names for the PDG codes of the particles are 
declared in \texttt{PDG.h}.  These are needed \eg to access the decay 
information.

\subsection{SLHA blocks}

The explicit indexing of the \texttt{slhadata} need not (and should not)
be done by the user.  Rather, the members of the SLHA data structure are
accessed through preprocessor variables.  Tables
\ref{tab:para1}--\ref{tab:para10} list the preprocessor variables
defined in \texttt{SLHA.h} which follow closely the definition of the
Accord.  Note that preprocessor symbols are case sensitive.  On the
downside, there is no way to guard against out-of-range indices, not
even with compiler flags.  This is because the preprocessor has no such
checks and the compiler cannot determine \textit{a posteriori} whether
the single index it sees addresses the `right' part of the array.

As far as there is overlap, the names for the block members have been
chosen similar to the ones used in the MSSM model file of
\textit{FeynArts} \cite{famssm}.  Following is a list of common index
conventions.  This is only for a rough orientation: the actual indices
and their ranges are always given explicitly in the Tables.
\begin{alignat*}{2}
\ran {t&} 1 4	& \qquad & \text{(s)fermion type:}
	\begin{tabular}[t]{l}
	1 = (s)neutrinos, \\
	2 = isospin-down (s)leptons, \\
	3 = isospin-up (s)quarks, \\
	4 = isospin-down (s)quarks
	\end{tabular} \\[1ex]
\ran {g&} 1 3	& & \text{(s)fermion generation} \\[1ex]
\ran {s&} 1 2	& & \text{number of sfermion mass-eigenstate,} \\
&		& & \qquad\text{in the absence of mixing 1 = L, 2 = R} \\[1ex]
\ran {c&} 1 2	& & \text{number of chargino mass-eigenstate} \\[1ex]
\ran {n&} 1 4	& & \text{number of neutralino mass-eigenstate}
\end{alignat*}
For each block $B$ the offset into \texttt{slhadata} and the length are
respectively defined as \texttt{Offset$B$} and \texttt{Length$B$}.  The 
contents of the block can be addressed through the macro 
\texttt{Block$B$($i$)}, where $i$ runs from 1 to \texttt{Length$B$}.

Matrices have a ``\texttt{Flat}'' array superimposed for convenience, in 
Fortran's standard column-major convention, \eg
\texttt{USf(1,1)} $\equiv$ \texttt{USfFlat(1)},
\texttt{USf(2,1)} $\equiv$ \texttt{USfFlat(2)}, 
\texttt{USf(1,2)} $\equiv$ \texttt{USfFlat(3)}, 
\texttt{USf(2,2)} $\equiv$ \texttt{USfFlat(4)}.
This makes it possible to \eg copy such a matrix with just a single
do-loop.

\begin{table}
\begin{slhablocks}
spinfo	& BlockSPInfo($n$) & SPInfo_Severity & \\
	& LengthSPInfo	& SPInfo_NLines & \\
	&		& SPInfo_Code($n$) & \ran n 1 {15} \\
	&		& SPInfo_Text($i$,$n$) & \textrm{do not address directly} \\
\hline
dcinfo	& BlockDCInfo($n$) & DCInfo_Severity & \\
	& LengthDCInfo	& DCInfo_NLines & \\
	&		& DCInfo_Code($n$) & \ran n 1 {15} \\
	&		& DCInfo_Text($i$,$n$) & \textrm{do not address directly} \\
\hline\hline
modsel	& BlockModSel($n$) & ModSel_Model & \\
	& LengthModSel	& ModSel_Content & \\
	&		& ModSel_RPV & \\
	&		& ModSel_CPV & \\
	&		& ModSel_FV & \\
	&		& ModSel_GridPts & \\
	&		& ModSel_Qmax & \\
	&		& ModSel_PDG($i$) & \ran i 1 5 \\
\hline\hline
sminputs & BlockSMInputs($n$) & SMInputs_AlfaMZ & \\
	& LengthSMInputs & SMInputs_invAlfaMZ & $\equiv$ SMInputs_AlfaMZ \\
	&		& SMInputs_GF	& \\
	&		& SMInputs_AlfasMZ & \\
	&		& SMInputs_MZ	& \\
	&		& SMInputs_Mf($t$,$g$) & \ran t 2 4,\ \ran g 1 3 \\
	&		& SMInputs_MfFlat($i$) & \ran i 1 9 \\
	&		& SMInputs_Me	& $\equiv$ SMInputs_Mf(2,1) \\
	&		& SMInputs_Mu	& $\equiv$ SMInputs_Mf(3,1) \\
	&		& SMInputs_Md	& $\equiv$ SMInputs_Mf(4,1) \\
	&		& SMInputs_Mmu	& $\equiv$ SMInputs_Mf(2,2) \\
	&		& SMInputs_Mc	& $\equiv$ SMInputs_Mf(3,2) \\
	&		& SMInputs_Ms	& $\equiv$ SMInputs_Mf(4,2) \\
	&		& SMInputs_Mtau	& $\equiv$ SMInputs_Mf(2,3) \\
	&		& SMInputs_Mt	& $\equiv$ SMInputs_Mf(3,3) \\
	&		& SMInputs_Mb	& $\equiv$ SMInputs_Mf(4,3) \\
\end{slhablocks}
\caption{\label{tab:para1}Preprocessor variables defined in
\texttt{SLHA.h} to access the \texttt{slhadata} array.
The equivalence symbol ($\equiv$) indicates that the l.h.s.\ is just an 
alias for the r.h.s., not a new variable.}
\end{table}

\begin{table}
\begin{slhablocks}
minpar	& BlockMinPar($n$) & MinPar_Q	& \\
	& LengthMinPar	& MinPar_M0	& \\
	&		& MinPar_Lambda	& $\equiv$ MinPar_M0 \\
	&		& MinPar_M12	& \\
	&		& MinPar_Mmess	& $\equiv$ MinPar_M12 \\
	&		& MinPar_M32	& $\equiv$ MinPar_M12 \\
	&		& MinPar_TB	& \\
	&		& MinPar_signMUE & \\
	&		& MinPar_A	& \\
	&		& MinPar_N5	& $\equiv$ MinPar_A \\
	&		& MinPar_cgrav	& \\
\hline\hline
extpar	& BlockExtPar($n$) & ExtPar_Q	& \\
	& LengthExtPar	& ExtPar_M1	& \\
	&		& ExtPar_M2	& \\
	&		& ExtPar_M3	& \\
	&		& ExtPar_Af($t$) & \ran t 2 4 \\
	&		& ExtPar_Atau	& $\equiv$ ExtPar_Af(2) \\
	&		& ExtPar_At	& $\equiv$ ExtPar_Af(3) \\
	&		& ExtPar_Ab	& $\equiv$ ExtPar_Af(4) \\
	&		& ExtPar_MHu2	& \\
	&		& ExtPar_MHd2	& \\
	&		& ExtPar_MUE	& \\
	&		& ExtPar_MA02	& \\
	&		& ExtPar_TB	& \\
	&		& ExtPar_MSS($g$,$q$) & \ran g 1 3,\ \ran q 1 5 \\
	&		& ExtPar_MSL($g$) & $\equiv$ ExtPar_MSS($g$,1) \\
	&		& ExtPar_MSE($g$) & $\equiv$ ExtPar_MSS($g$,2) \\
	&		& ExtPar_MSQ($g$) & $\equiv$ ExtPar_MSS($g$,3) \\
	&		& ExtPar_MSU($g$) & $\equiv$ ExtPar_MSS($g$,4) \\
	&		& ExtPar_MSD($g$) & $\equiv$ ExtPar_MSS($g$,5) \\
	&		& ExtPar_N5($g$) & \ran g 1 3 \\
	&		& ExtPar_lambda	& \\
	&		& ExtPar_kappa	& \\
	&		& ExtPar_Alambda & \\
	&		& ExtPar_Akappa	& \\
	&		& ExtPar_MUEeff	& \\
\end{slhablocks}
\caption{\label{tab:para2}Preprocessor variables defined in
\texttt{SLHA.h} to access the \texttt{slhadata} array (cont'd).}
\end{table}

\begin{table}
\begin{slhablocks}
mass	& BlockMass($n$) & Mass_Mf($t$,$g$) & \ran t 1 4,\ \ran g 1 3 \\
	& LengthMass	& Mass_MfFlat($i$) & \ran i 1 {12} \\
	&		& Mass_MSf($s$,$t$,$g$) & \ran s 1 2,\ \ran t 1 4,\ \ran g 1 3 \\
	&		& Mass_MSfFlat($i$) & \ran i 1 {24} \\
	&		& Mass_MZ	& \\
	&		& Mass_MW	& \\
	&		& Mass_Mh0	& \\
	&		& Mass_MHH	& \\
	&		& Mass_MA0	& \\
	&		& Mass_MHp	& \\
	&		& Mass_MH1	& $\equiv$ Mass_Mh0 \\
	&		& Mass_MH2	& $\equiv$ Mass_MHH \\
	&		& Mass_MH3	& \\
	&		& Mass_MA1	& $\equiv$ Mass_MA0 \\
	&		& Mass_MA2	& \\
	&		& Mass_MNeu($n$) & \ran n 1 5 \\
	&		& Mass_MCha($c$) & \ran c 1 2 \\
	&		& Mass_MGl	& \\
	&		& Mass_MGrav	& \\
\hline\hline
nmix	& BlockNMix($n$) & NMix_ZNeu($n!1$,$n!2$) & \ran {n!1,n!2} 1 4 \\
	& LengthNMix	& NMix_ZNeuFlat($i$)	& \ran i 1 {16} \\
\hline\hline
umix	& BlockUMix($n$) & UMix_UCha($c!1$,$c!2$) & \ran {c!1,c!2} 1 2 \\
	& LengthUMix	& UMix_UChaFlat($i$)	& \ran i 1 4 \\
\hline
vmix	& BlockVMix($n$) & VMix_VCha($c!1$,$c!2$) & \ran {c!1,c!2} 1 2 \\
	& LengthVMix	& VMix_VChaFlat($i$)	& \ran i 1 4 \\
\end{slhablocks}
\caption{\label{tab:para3}Preprocessor variables defined in
\texttt{SLHA.h} to access the \texttt{slhadata} array (cont'd).}
\end{table}

\begin{table}
\begin{slhablocks}
	& BlockSfMix($n$) & SfMix_USf($s!1$,$s!2$,$t$) & \ran {s!1,s!2} 1 2,\ \ran t 2 4 \\
	& LengthSfMix	& SfMix_USfFlat($i$,$t$) & \ran i 1 4,\ \ran t 2 4 \\
\hline
staumix	& BlockStauMix($n$) & StauMix_USf($s!1$,$s!2$) & $\equiv$ SfMix_USf($s!1$,$s!2$,2) \\
	& LengthStauMix	& StauMix_USfFlat($i$)	& $\equiv$ SfMix_USfFlat($i$,2) \\
\hline
stopmix	& BlockStopMix($n$) & StopMix_USf($s!1$,$s!2$) & $\equiv$ SfMix_USf($s!1$,$s!2$,3) \\
	& LengthStopMix	& StopMix_USfFlat($i$)	& $\equiv$ SfMix_USfFlat($i$,3) \\
\hline
sbotmix	& BlockSbotMix($n$) & SbotMix_USf($s!1$,$s!2$) & $\equiv$ SfMix_USf($s!1$,$s!2$,4) \\
	& LengthSbotMix	& SbotMix_USfFlat($i$)	& $\equiv$ SfMix_USfFlat($i$,4) \\
\hline\hline
alpha	& BlockAlpha($n$) & Alpha_Alpha	& \\
	& LengthAlpha	& 		& \\
\hline\hline
hmix	& BlockHMix($n$) & HMix_Q	& \\
	& LengthHMix	& HMix_MUE	& \\
	&		& HMix_TB	& \\
	&		& HMix_VEV	& \\
	&		& HMix_MA02	& \\
\hline\hline
gauge	& BlocktGauge($n$) & Gauge_Q	& \\
	& LengthGauge	& Gauge_g1	& \\
	&		& Gauge_g2	& \\
	&		& Gauge_g3	& \\
\hline\hline
msoft	& BlockMSoft($n$) & MSoft_Q	& \\
	& LengthMSoft	& MSoft_M1	& \\
	&		& MSoft_M2	& \\
	&		& MSoft_M3	& \\
	&		& MSoft_MHu2	& \\
	&		& MSoft_MHd2	& \\
	&		& MSoft_MSS($g$,$q$) & \ran g 1 3,\ \ran q 1 5 \\
	&		& MSoft_MSL($g$) & $\equiv$ MSoft_MSS($g$,1) \\
	&		& MSoft_MSE($g$) & $\equiv$ MSoft_MSS($g$,2) \\
	&		& MSoft_MSQ($g$) & $\equiv$ MSoft_MSS($g$,3) \\
	&		& MSoft_MSU($g$) & $\equiv$ MSoft_MSS($g$,4) \\
	&		& MSoft_MSD($g$) & $\equiv$ MSoft_MSS($g$,5) \\
\end{slhablocks}
\caption{\label{tab:para4}Preprocessor variables defined in
\texttt{SLHA.h} to access the \texttt{slhadata} array (cont'd).}
\end{table}

\begin{table}
\begin{slhablocks}
	& BlockAf($n$)	& Af_Q($t$)	& \ran t 2 4 \\
	& LengthAf	& Af_Af($g!1$,$g!2$,$t$) & \ran {g!1,g!2} 1 3,\ \ran t 2 4 \\
	&		& Af_AfFlat($i$,$t$) & \ran i 1 9,\ \ran t 2 4 \\
\hline
ae	& BlockAe($n$)	& Ae_Q		& $\equiv$ Af_Q(2) \\
	& LengthAe	& Ae_Af($g!1$,$g!2$) & $\equiv$ Af_Af($g!1$,$g!2$,2) \\
	&		& Ae_AfFlat($i$) & $\equiv$ Af_AfFlat($i$,2) \\
	&		& Ae_Atau	& $\equiv$ Ae_Af(3,3) \\
\hline
au	& BlockAu($n$)	& Au_Q		& $\equiv$ Af_Q(3) \\
	& LengthAu	& Au_Af($g!1$,$g!2$) & $\equiv$ Af_Af($g!1$,$g!2$,3) \\
	&		& Au_AfFlat($i$) & $\equiv$ Af_AfFlat($i$,3) \\
	&		& Au_At		& $\equiv$ Au_Af(3,3) \\
\hline
ad	& BlockAd($n$)	& Ad_Q		& $\equiv$ Af_Q(4) \\
	& LengthAd	& Ad_Af($g!1$,$g!2$) & $\equiv$ Af_Af($g!1$,$g!2$,4) \\
	&		& Ad_AfFlat($i$) & $\equiv$ Af_AfFlat($i$,4) \\
 	&		& Ad_Ab		& $\equiv$ Ad_Af(3,3) \\
\hline\hline
	& BlockYf($n$)	& Yf_Q($t$)	& \ran t 2 4 \\
	& LengthYf	& Yf_Af($g!1$,$g!2$,$t$) & \ran {g!1,g!2} 1 3,\ \ran t 2 4 \\
	&		& Yf_AfFlat($i$,$t$) & \ran i 1 9,\ \ran t 2 4 \\
\hline
ye	& BlockYe($n$)	& Ye_Q		& $\equiv$ Yf_Q(2) \\
	& LengthYe	& Ye_Yf($g!1$,$g!2$) & $\equiv$ Yf_Yf($g!1$,$g!2$,2) \\
	&		& Ye_YfFlat($i$) & $\equiv$ Yf_YfFlat($i$,2) \\
	&		& Ye_Atau	& $\equiv$ Ye_Yf(3,3) \\
\hline
yu	& BlockYu($n$)	& Yu_Q		& $\equiv$ Yf_Q(3) \\
	& LengthYu	& Yu_Yf($g!1$,$g!2$) & $\equiv$ Yf_Yf($g!1$,$g!2$,3) \\
	&		& Yu_YfFlat($i$) & $\equiv$ Yf_YfFlat($i$,3) \\
	&		& Yu_At		& $\equiv$ Yu_Yf(3,3) \\
\hline
yd	& BlockYd($n$)	& Yd_Q		& $\equiv$ Yf_Q(4) \\
	& LengthYd	& Yd_Yf($g!1$,$g!2$) & $\equiv$ Yf_Yf($g!1$,$g!2$,4) \\
	&		& Yd_YfFlat($i$) & $\equiv$ Yf_YfFlat($i$,4) \\
	&		& Yd_Ab		& $\equiv$ Yd_Yf(3,3) \\
\end{slhablocks}
\caption{\label{tab:para5}Preprocessor variables defined in
\texttt{SLHA.h} to access the \texttt{slhadata} array (cont'd).}
\end{table}

\begin{table}
\begin{slhablocks}
rvlambdain & BlockRVLambdaIn($n$) & RVLambdaIn_lambda($i$,$j$,$k$) & \ran {i,j,k} 1 3 \\
	& LengthRVLambdaIn & RVLambdaIn_lambdaFlat($i$) & \ran i 1 {27} \\
\hline
rvlambda & BlockRVLambda($n$) & RVLambda_Q & \\
	& LengthRVLambda & RVLambda_lambda($i$,$j$,$k$) & \ran {i,j,k} 1 3 \\
	&		& RVLambda_lambdaFlat($i$) & \ran i 1 {27} \\
\hline\hline
rvlambdapin & BlockRVLambdaPIn($n$) & RVLambdaPIn_lambdaP($i$,$j$,$k$) & \ran {i,j,k} 1 3 \\
	& LengthRVLambdaPIn & RVLambdaPIn_lambdaPFlat($i$) & \ran i 1 {27} \\
\hline
rvlambdap & BlockRVLambdaP($n$) & RVLambdaP_Q & \\
	& LengthRVLambdaP & RVLambdaP_lambdaP($i$,$j$,$k$) & \ran {i,j,k} 1 3 \\
	&		& RVLambdaP_lambdaPFlat($i$) & \ran i 1 {27} \\
\hline\hline
rvlambdappin & BlockRVLambdaPPIn($n$) & RVLambdaPPIn_lambdaPP($i$,$j$,$k$) & \ran {i,j,k} 1 3 \\
	& LengthRVLambdaPPIn & RVLambdaPPIn_lambdaPPFlat($i$) & \ran i 1 {27} \\
\hline
rvlambdapp & BlockRVLambdaPP($n$) & RVLambdaPP_Q & \\
	& LengthRVLambdaPP & RVLambdaPP_lambdaPP($i$,$j$,$k$) & \ran {i,j,k} 1 3 \\
	&		& RVLambdaPP_lambdaPPFlat($i$) & \ran i 1 {27} \\
\hline\hline
rvain	& BlockRVAIn($n$) & RVAIn_A($i$,$j$,$k$) & \ran {i,j,k} 1 3 \\
	& LengthRVAIn	& RVAIn_AFlat($i$) & \ran i 1 {27} \\
\hline
rva	& BlockRVA($n$)	& RVA_Q		& \\
	& LengthRVA	& RVA_A($i$,$j$,$k$) & \ran {i,j,k} 1 3 \\
	&		& RVA_AFlat($i$) & \ran i 1 {27} \\
\hline\hline
rvapin	& BlockRVAPIn($n$) & RVAPIn_AP($i$,$j$,$k$) & \ran {i,j,k} 1 3 \\
	& LengthRVAPIn	& RVAPIn_APFlat($i$) & \ran i 1 {27} \\
\hline
rvap	& BlockRVAP($n$) & RVAP_Q	& \\
	& LengthRVAP	& RVAP_AP($i$,$j$,$k$) & \ran {i,j,k} 1 3 \\
	&		& RVAP_APFlat($i$) & \ran i 1 {27} \\
\hline\hline
rvappin & BlockRVAPPIn($n$) & RVAPPIn_APP($i$,$j$,$k$) & \ran {i,j,k} 1 3 \\
	& LengthRVAPPIn	& RVAPPIn_APPFlat($i$) & \ran i 1 {27} \\
\hline
rvapp	& BlockRVAPP	& RVAPP_Q	& \\
	& LengthRVAPP	& RVAPP_APP($i$,$j$,$k$) & \ran {i,j,k} 1 3 \\
	&		& RVAPP_APPFlat($i$) & \ran i 1 {27} \\
\end{slhablocks}
\caption{\label{tab:para6}Preprocessor variables defined in
\texttt{SLHA.h} to access the \texttt{slhadata} array (cont'd).}
\end{table}

\begin{table}
\begin{slhablocks}
rvkappain & BlockRVKappaIn($n$) & RVKappaIn_kappa($i$) & \ran i 1 3 \\
	& LengthRVKappaIn &		& \\
\hline
rvkappa & BlockRVKappa($n$) & RVKappa_Q & \\
	& LengthRVKappa & RVKappa_kappa($i$) & \ran i 1 3 \\
\hline\hline
rvdin	& BlockRVDIn($n$) & RVDIn_D($i$) & \ran i 1 3 \\
	& LengthRVDIn	&		& \\
\hline
rvd	& BlockRVD($n$)	& RVD_Q 	& \\
	& LengthRVD	& RVD_D($i$)	& \ran i 1 3 \\
\hline\hline
rvsnvevin & BlockRVSnVEVIn($n$) & RVSnVEVIn_VEV($i$) & \ran i 1 3 \\
	& LengthRVSnVEVIn &		& \\
\hline
rvsnvev	& BlockRVSnVEV($n$) & RVSnVEV_Q & \\
	& LengthRVSnVEV	& RVSnVEV_VEV($i$) & \ran i 1 3 \\
\hline\hline
rvmlh1sqin & BlockRVMLH1SqIn($n$) & RVMLH1SqIn_MLH12($i$) & \ran i 1 3 \\
	& LengthRVMLH1SqIn &		& \\ \hline
rvmlh1sq & BlockRVMLH1Sq($n$) & RVMLH1Sq_Q & \\
	& LengthRVMLH1Sq & RVMLH1Sq_MLH12($i$) & \ran i 1 3 \\
\hline\hline
rvnmix	& BlockRVNMix($n$) & RVNMix_ZNeu($n!1$,$n!2$) & \ran {n!1,n!2} 1 7 \\
	& LengthRVNMix	& RVNMix_ZNeuFlat($i$)	& \ran i 1 {49} \\
\hline\hline
rvumix	& BlockRVUMix($n$) & RVUMix_UCha($c!1$,$c!2$) & \ran {c!1,c!2} 1 5 \\
	& LengthRVUMix	& RVUMix_UChaFlat($i$)	& \ran i 1 {25} \\
\hline
rvvmix	& BlockRVVMix($n$) & RVVMix_VCha($c!1$,$c!2$) & \ran {c!1,c!2} 1 5 \\
	& LengthRVVMix	& RVVMix_VChaFlat($i$)	& \ran i 1 {25} \\
\hline\hline
rvhmix	& BlockRVHMix($n$) & RVUMix_UH($h!1$,$h!2$) & \ran {h!1,h!2} 1 5 \\
	& LengthRVHMix	& RVUMix_UHFlat($i$)	& \ran i 1 {25} \\
\hline
rvamix	& BlockRVAMix($n$) & RVAMix_UA($h!1$,$h!2$) & \ran {h!1,h!2} 1 5 \\
	& LengthRVAMix	& RVAMix_UAFlat($i$)	& \ran i 1 {25} \\
\hline\hline
rvlmix	& BlockRVLMix($n$) & RVLMix_CLep($l!1$,$l!2$) & \ran {l!1,l!2} 1 8 \\
	& LengthRVLMix	& RVLMix_CLepFlat($i$)	& \ran i 1 {64} \\
\end{slhablocks}
\caption{\label{tab:para7}Preprocessor variables defined in
\texttt{SLHA.h} to access the \texttt{slhadata} array (cont'd).}
\end{table}

\begin{table}
\begin{slhablocks}
vckminputs & BlockVCKMInputs($n$) & VCKMInputs_theta12 & \\
	& LengthVCKMInputs & VCKMInputs_theta23	& \\
	&		& VCKMInputs_theta13 	& \\
	&		& VCKMInputs_delta13 	& \\
\hline
vckm	& BlockVCKM($n$) & VCKM_Q		& \\
	& LengthVCKM	& VCKM_theta12		& \\
	&		& VCKM_theta23		& \\
	&		& VCKM_theta13		& \\
	&		& VCKM_delta13		& \\
\end{slhablocks}
\caption{\label{tab:para8}Preprocessor variables defined in
\texttt{SLHA.h} to access the \texttt{slhadata} array (cont'd).}
\end{table}

\begin{table}
\begin{slhablocks}
      	& BlockMSS2In($n$) & MSS2In_MSS2($g!1$,$g!2$,$q$) & \ran {g!1,g!2} 1 3,\ \ran q 1 5 \\
	& LengthMSS2In	& MSS2In_MSS2Flat($i$,$q$) & \ran i 1 9,\ \ran q 1 5 \\
\hline
msl2in	& BlockMSL2In($n$) & MSL2In_MSL2($g!1$,$g!2$) & $\equiv$ MSS2In_MSS2($g!1$,$g!2$,1) \\
	& LengthMSL2In	& MSL2In_MSL2Flat($i$)	& $\equiv$ MSS2In_MSS2Flat($i$,1) \\
\hline
mse2in	& BlockMSE2In($n$) & MSE2In_MSE2($g!1$,$g!2$) & $\equiv$ MSS2In_MSS2($g!1$,$g!2$,2) \\
	& LengthMSE2In	& MSE2In_MSE2Flat($i$)	& $\equiv$ MSS2In_MSS2Flat($i$,2) \\
\hline
msq2in	& BlockMSQ2In($n$) & MSQ2In_MSQ2($g!1$,$g!2$) & $\equiv$ MSS2In_MSS2($g!1$,$g!2$,3) \\
	& LengthMSQ2In	& MSQ2In_MSQ2Flat($i$)	& $\equiv$ MSS2In_MSS2Flat($i$,3) \\
\hline
msu2in	& BlockMSU2In($n$) & MSU2In_MSU2($g!1$,$g!2$) & $\equiv$ MSS2In_MSS2($g!1$,$g!2$,4) \\
	& LengthMSU2In	& MSU2In_MSU2Flat($i$)	& $\equiv$ MSS2In_MSS2Flat($i$,4) \\
\hline
msd2in	& BlockMSD2In($n$) & MSD2In_MSD2($g!1$,$g!2$) & $\equiv$ MSS2In_MSS2($g!1$,$g!2$,5) \\
	& LengthMSD2In	& MSD2In_MSD2Flat($i$)	& $\equiv$ MSS2In_MSS2Flat($i$,5) \\
\hline\hline
      	& BlockMSS2($n$) & MSS2_Q($q$)		& \ran q 1 5 \\
	& LengthMSS2	& MSS2_MSS2($g!1$,$g!2$,$q$) & \ran {g!1,g!2} 1 3,\ \ran q 1 5 \\
	&		& MSS2_MSS2Flat($i$,$q$) & \ran i 1 9,\ \ran q 1 5 \\
\hline
msl2	& BlockMSL2($n$) & MSL2_Q		& $\equiv$ MSS2_Q(1) \\
	& LengthMSL2	& MSL2_MSL2($g!1$,$g!2$) & $\equiv$ MSS2_MSS2($g!1$,$g!2$,1) \\
	&		& MSL2_MSL2Flat($i$)	& $\equiv$ MSS2_MSS2Flat($i$,1) \\
\hline
mse2	& BlockMSE2($n$) & MSE2_Q		& $\equiv$ MSS2_Q(2) \\
	& LengthMSE2	& MSE2_MSE2($g!1$,$g!2$) & $\equiv$ MSS2_MSS2($g!1$,$g!2$,2) \\
	&		& MSE2_MSE2Flat($i$)	& $\equiv$ MSS2_MSS2Flat($i$,2) \\
\hline
msq2	& BlockMSQ2($n$) & MSQ2_Q		& $\equiv$ MSS2_Q(3) \\
	& LengthMSQ2	& MSQ2_MSQ2($g!1$,$g!2$) & $\equiv$ MSS2_MSS2($g!1$,$g!2$,3) \\
	&		& MSQ2_MSQ2Flat($i$)	& $\equiv$ MSS2_MSS2Flat($i$,3) \\
\hline
msu2	& BlockMSU2($n$) & MSU2_Q		& $\equiv$ MSS2_Q(4) \\
	& LengthMSU2	& MSU2_MSU2($g!1$,$g!2$) & $\equiv$ MSS2_MSS2($g!1$,$g!2$,4) \\
	&		& MSU2_MSU2Flat($i$)	& $\equiv$ MSS2_MSS2Flat($i$,4) \\
\hline
msd2	& BlockMSD2($n$) & MSD2_Q		& $\equiv$ MSS2_Q(5) \\
	& LengthMSD2	& MSD2_MSD2($g!1$,$g!2$) & $\equiv$ MSS2_MSS2($g!1$,$g!2$,5) \\
	&		& MSD2_MSD2Flat($i$)	& $\equiv$ MSS2_MSS2Flat($i$,5) \\
\hline\hline
	& BlockASfMix($n$) & ASfMix_UASf($s!1$,$s!2$,$t$) & \ran {s!1,s!2} 1 6,\ \ran t 1 4 \\
	& LengthASfMix	& ASfMix_UASfFlat($i$,$t$)	& \ran i 1 {36},\ \ran t 1 4 \\
\hline
snmix	& BlockSnMix($n$) & SnMix_UASf($s!1$,$s!2$) & $\equiv$ ASfMix_UASf($s!1$,$s!2$,1) \\
	& LengthSnMix	& SnMix_UASfFlat($i$)	& $\equiv$ ASfMix_UASfFlat($i$,1) \\
\hline
slmix	& BlockSlMix($n$) & SlMix_UASf($s!1$,$s!2$) & $\equiv$ ASfMix_UASf($s!1$,$s!2$,2) \\
	& LengthSlMix	& SlMix_UASfFlat($i$)	& $\equiv$ ASfMix_UASfFlat($i$,2) \\
\hline
usqmix	& BlockUSqMix($n$) & USqMix_UASf($s!1$,$s!2$) & $\equiv$ ASfMix_UASf($s!1$,$s!2$,3) \\
	& LengthUSqMix	& USqMix_UASfFlat($i$)	& $\equiv$ ASfMix_UASfFlat($i$,3) \\
\hline
dsqmix	& BlockDSqMix($n$) & DSqMix_UASf($s!1$,$s!2$) & $\equiv$ ASfMix_UASf($s!1$,$s!2$,4) \\
	& LengthDSqMix	& DSqMix_UASfFlat($i$)	& $\equiv$ ASfMix_UASfFlat($i$,4) \\
\end{slhablocks}
\caption{\label{tab:para9}Preprocessor variables defined in
\texttt{SLHA.h} to access the \texttt{slhadata} array (cont'd).}
\end{table}

\begin{table}
\begin{slhablocks}
cvhmix	& BlockCVHMix($n$) & CVHMix_UH($h!1$,$h!2$) & \ran {h!1,h!2} 1 4 \\
	& LengthCVHMix	& CVHMix_UHFlat($i$)	& \ran i 1 {16} \\
\hline\hline
nmnmix	& BlockNMNMix($n$) & NMNMix_ZNeu($n!1$,$n!2$) & \ran {n!1,n!2} 1 5 \\
	& LengthNMNMix	& NMNMix_ZNeuFlat($i$)	& \ran i 1 {25} \\
\hline\hline
nmhmix	& BlockNMHMix($n$) & NMUMix_UH($h!1$,$h!2$) & \ran {h!1,h!2} 1 3 \\
	& LengthNMHMix	& NMUMix_UHFlat($i$)	& \ran i 1 9 \\
\hline
nmamix	& BlockNMAMix	& NMAMix_UA($h!1$,$h!2$) & \ran {h!1,h!2} 1 3 \\
	& LengthNMAMix	& NMAMix_UAFlat($i$)	& \ran i 1 9 \\
\end{slhablocks}
\caption{\label{tab:para10}Preprocessor variables defined in
\texttt{SLHA.h} to access the \texttt{slhadata} array (cont'd).}
\end{table}

\subsection{PDG particle identifiers}
\label{sect:pdg}

\texttt{PDG.h} defines the human-readable versions of the PDG codes
listed in Table~\ref{tab:pdg}.  These are needed \eg to access the decay
information.  At run time, the subroutine \texttt{SLHAPDGName} can be
used to translate a PDG code into a particle name (see 
Sect.~\ref{sect:pdgname}).

\begin{table}
\begin{center}
\begin{small}
\begin{tt}
\begin{tabular}[t]{|l|ll|l|} \hline
\textnormal{fermions} &
\textnormal{sfermions} & \\
\hline\hline
PDG_nu_e	& PDG_snu_e1	& PDG_snu_e2 \\
PDG_electron	& PDG_selectron1 & PDG_selectron2 \\
PDG_up		& PDG_sup1	& PDG_sup2 \\
PDG_down	& PDG_sdown1	& PDG_sdown2 \\
\hline
PDG_nu_mu	& PDG_snu_mu1	& PDG_snu_mu2 \\
PDG_muon	& PDG_smuon1	& PDG_smuon2 \\
PDG_charm	& PDG_scharm1	& PDG_scharm2 \\
PDG_strange	& PDG_sstrange1	& PDG_sstrange2 \\
\hline
PDG_nu_tau	& PDG_snu_tau1	& PDG_snu_tau2 \\
PDG_tau		& PDG_stau1	& PDG_stau2 \\
PDG_top		& PDG_stop1	& PDG_stop2 \\
PDG_bottom	& PDG_sbottom1	& PDG_sbottom2 \\
\hline
\end{tabular}
\begin{tabular}[t]{|l|l|} \hline
\textnormal{bosons} &
\textnormal{gauginos} \\
\hline\hline
PDG_h0		& PDG_neutralino1 \\
PDG_HH		& PDG_neutralino2 \\
PDG_A0		& PDG_neutralino3 \\
PDG_Hp		& PDG_neutralino4 \\
PDG_H3		& PDG_neutralino5 \\
PDG_A2		& PDG_chargino1 \\
PDG_photon	& PDG_chargino2 \\
PDG_Z		& PDG_gluino \\
PDG_W		& PDG_gravitino \\
PDG_gluon	& \\
PDG_graviton	& \\
\hline
\end{tabular}
\end{tt}
\end{small}
\end{center}
\caption{\label{tab:pdg}The PDG codes defined in \texttt{PDG.h}.}
\end{table}


\newpage

\section{Routines provided by the SLHA library}
\label{sect:ref}

The file \texttt{SLHA.h} must be included in every subroutine or
function that uses SLHALib routines.  It contains the necessary
preprocessor definitions as well as external declarations for the
SLHALib routines.

The basic data structure is the double complex array \texttt{slhadata}
of length \texttt{nslhadata}.  These names are hard-coded into the
preprocessor definitions and may not be changed by the user.  As a
corollary, only one instance of the \texttt{slhadata} structure can be
used in any one routine.  This poses no serious limitation for most 
applications, however.

\subsection{SLHAClear}

\begin{verbatim}
        subroutine SLHAClear(slhadata)
        double complex slhadata(nslhadata)
\end{verbatim}
This subroutine sets all data in the \texttt{slhadata} array given as
argument to the value \texttt{invalid} (defined in \texttt{SLHA.h}).  It
is important that this is done before using \texttt{slhadata}, or else
any kind of junk that happens to be in the memory occupied by
\texttt{slhadata} will later on be interpreted as valid data.

\subsection{SLHARead}

\begin{verbatim}
        subroutine SLHARead(error, slhadata, filename, abort)
        integer error, abort
        double complex slhadata(nslhadata)
        character*(*) filename
\end{verbatim}
This subroutine reads the data in SLHA format from \texttt{filename}
into the \texttt{slhadata} array.  If the specified file cannot be
opened, the function issues an error message and returns \texttt{error =
1}.  The \texttt{abort} flag governs what happens when superfluous text
is read, \ie text that cannot be interpreted as SLHA data.  If
\texttt{abort} is 0, a warning is printed and reading continues. 
Otherwise, reading stops at the offending line and \texttt{error = 2} is
returned.  \texttt{SLHARead} implicitly calls \texttt{SLHAClear} to
clear the \texttt{slhadata} array before reading the file.

The blocks \textsc{spinfo} and \textsc{dcinfo} are largely ignored when
reading the file, as they are for human information only.  Only the
maximum of all message codes is kept in the \texttt{Severity} member of
the block.  Since the message codes increase with severity, this
indicates the overall reliability of the corresponding data (spectrum or
decay information).  For example, if the \texttt{Severity} member is 4
(real errors), the Accord advises not to use the corresponding data. 
See also Sect.~\ref{sect:info}.

\subsection{SLHAWrite}

{\samepage
\begin{verbatim}
        subroutine SLHAWrite(error, slhadata, filename)
        integer error
        double complex slhadata(nslhadata)
        character*(*) filename
\end{verbatim}}
This subroutine writes the data in \texttt{slhadata} to
\texttt{filename}.

\subsection{SLHAInfo}
\label{sect:info}

\begin{verbatim}
        subroutine SLHAInfo(slhablock, code, text)
        double complex slhablock(*)
        integer code
        character*(*) text
\end{verbatim}
This subroutine adds a message to one of the informational blocks, 
\textsc{spinfo} or \textsc{dcinfo}.  The block is most conveniently
addressed through the \texttt{Block...} macros, for example
\begin{verbatim}
   call SLHAInfo(BlockSPInfo(1), 4, "Error in computation")
\end{verbatim}
Allowed codes are
\begin{itemize}
\item 1 = program name,
\item 2 = program version,
\item 3 = warning message,
\item 4 = error message.
\end{itemize}
Messages are truncated at 80 characters.

\subsection{SLHANewDecay}

\begin{verbatim}
        integer function SLHANewDecay(slhadata, width, parent_id)
        double complex slhadata(nslhadata)
        double precision width
        integer parent_id
\end{verbatim}
This function initiates the setting of decay information for the
particle specified by the \texttt{parent_id} PDG code, whose total decay
width is given by \texttt{width}.  The return value is an integer index 
which is needed to subsequently add individual decay modes with
\texttt{SLHAAddDecay}.  If the fixed-length array \texttt{slhadata} 
becomes full, a warning is printed and zero is returned.  If a decay of 
the given particle is already present in \texttt{slhadata}, it is first 
removed.

\subsection{SLHAFindDecay}

\begin{verbatim}
        integer function SLHAFindDecay(slhadata, parent_id)
        double complex slhadata(nslhadata)
        integer parent_id
\end{verbatim}
This function also initiates the setting of decay information.  Unlike
\texttt{SLHANewDecay}, it requires that the decay of the
\texttt{parent_id} particle exist and reshuffles the decay information
in \texttt{slhadata} such that new channels can be added to this
decay.  If no decay matching \texttt{parent_id} is found, the return
value is 0, otherwise it is the index needed to add decay modes with
\texttt{SLHAAddDecay}.

\subsection{SLHAAddDecay}

\begin{verbatim}
        subroutine SLHAAddDecay(slhadata, br, decay,
     &    nchildren, child1_id, child2_id, child3_id, child4_id)
        double complex slhadata(nslhadata)
        double precision br
        integer decay
        integer nchildren, child1_id, child2_id, child3_id, child4_id
\end{verbatim}
This subroutine adds the decay mode
$$
\texttt{(parent_id)}\quad\to
\quad\texttt{child1_id}
\quad\texttt{child2_id}
\quad\texttt{child3_id}
\quad\texttt{child4_id}
$$
to the decay section previously initiated by \texttt{SLHANewDecay} or
\texttt{SLHAFindDecay}.  \texttt{decay} is the index obtained from the
latter (which also set the \texttt{parent_id}) and \texttt{child$n$_id}
are the PDG codes of the final-state particles.  The branching ratio is
given in \texttt{br}.  If the fixed-length array \texttt{slhadata}
becomes full, a warning is printed and \texttt{decay} is set to zero.

If \texttt{decay} is zero, an overflow of \texttt{slhadata} in an
earlier invocation is silently assumed and no action is performed.  It
is therefore sufficient to check for overflow only once, after setting
all decay modes (unless, of course, one needs to pinpoint the exact
location of the overflow).

As with \texttt{SLHAGetDecay} (see Sect.~\ref{sect:getdecay}), only the
first \texttt{nchildren} of the \texttt{child$n$_id} are actually
accessed and Fortran allows to omit the remaining ones in the
invocation.

\subsection{SLHAGetDecay}
\label{sect:getdecay}

\begin{verbatim}
        double precision function SLHAGetDecay(slhadata, parent_id,
     &    nchildren, child1_id, child2_id, child3_id, child4_id)
        double complex slhadata(*)
        integer parent_id
        integer nchildren, child1_id, child2_id, child3_id, child4_id
\end{verbatim}
This function extracts the decay
$$
\texttt{parent_id}\quad\to
\quad\texttt{child1_id}
\quad\texttt{child2_id}
\quad\texttt{child3_id}
\quad\texttt{child4_id}
$$
from the \texttt{slhadata} array, or the value \texttt{invalid} (defined
in \texttt{SLHA.h}) if no such decay can be found.  The parent and child
particles are given by their PDG identifiers (see Sect.~\ref{sect:pdg}).  
The return value is the total decay width if \texttt{nchildren = 0}, 
otherwise the branching ratio of the specified channel.

Note that only the first \texttt{nchildren} of the \texttt{child$n$_id}
are actually accessed and Fortran allows to omit the remaining ones in
the invocation (a strict syntax checker might issue a warning, though). 
Thus, for instance,
\begin{verbatim}
   Zbb = SLHAGetDecay(slhadata, PDG_Z, 2, PDG_bottom, -PDG_bottom)
\end{verbatim}
is a perfectly legitimate way to extract the $Z\to b\bar b$ decay.

\subsection{SLHADecayTable}

\begin{verbatim}
        integer function SLHADecayTable(slhadata, parent_id,
     &    width, id, maxparticles, maxchannels)
        double complex slhadata(nslhadata)
        integer parent_id, maxparticles, maxchannels
        double precision width(maxchannels)
        integer id(0:maxparticles,maxchannels)
\end{verbatim}
This function stores all decay channels for the particle identified by
\texttt{parent_id} in the arrays \texttt{id} and \texttt{width}.  Unlike
\texttt{SLHAGetDecay}, one does not need to know the exact decay mode in
order to extract information.  The value 0 for \texttt{parent_id} serves
as a wildcard and transfers the entire decay table contained in
\texttt{slhadata}.  \texttt{SLHADecayTable} returns the number of
channels found.  The two arrays can be read out rather
straightforwardly:

For each channel $c$,
\begin{itemize}
\item $n$ = \texttt{id(0,$c$)} gives the number of participating 
      particles, \ie the number of decay products plus one.
\item The PDG code of the decaying particle is in \texttt{id(1,$c$)}.
\item The PDG codes of the decay products are in \texttt{id(2,$c$)}\dots
      \texttt{id($n$,$c$)}.
\item If $n = 1$, \texttt{width($c$)} contains the decaying particle's 
      total width in GeV.
\item If $n > 1$, \texttt{width($c$)} contains the branching ratio
      for the given decay.
\end{itemize}

\subsection{SLHAExist}

\begin{verbatim}
        integer function SLHAExist(slhablock, length)
        double complex slhablock(*)
        integer length
\end{verbatim}
This function tests whether a given SLHA block is not entirely empty. It
returns 2 if the block has at least one complex member, 1 if the block
has at least one real member (\ie all imaginary parts zero), and 0 if
the block has no valid members at all.  The SLHA blocks are most 
conveniently accessed using the \texttt{Block...} and \texttt{Length...} 
definitions (see Sect.~\ref{sect:data}), \eg
\begin{verbatim}
        if( SLHAExist(BlockMass(1), LengthMass) .ne. 0 ) ...
\end{verbatim}

\subsection{SLHAValid}

\begin{verbatim}
        logical function SLHAValid(slhablock, length)
        double complex slhablock(*)
        integer length
\end{verbatim}
This function tests whether a given SLHA block consists entirely of
valid data, \ie it returns \texttt{.FALSE.} if at least one member of
the block is invalid.  The SLHA blocks are most conveniently accessed
using the \texttt{Block...}  and \texttt{Length...} definitions (see
Sect.~\ref{sect:data}), \eg
\begin{verbatim}
        if( SLHAValid(BlockNMix(1), LengthNMix) ) ...
\end{verbatim}

\subsection{SLHAPDGName}
\label{sect:pdgname}

\begin{verbatim}
        subroutine SLHAPDGName(code, name)
        integer code
        character*(PDGLen) name
\end{verbatim}
This subroutine translates a PDG code into a particle name.  The sign of 
the PDG code is ignored, hence the same name is returned for a particle 
and its antiparticle.  The maximum length of the name, \texttt{PDGLen}, 
is defined in \texttt{PDG.h}.

\subsection{Incompatible Changes}
\label{sect:compat}

Two incompatible changes in the interface were necessary with respect 
to the SLHALib 1 \cite{slhalib1}, largely due to the fact that the SLHA2 
allows complex entries:
\begin{itemize}
\item \texttt{slhadata} is now a double complex, not a double
      precision array.
\item The \texttt{SLHAExist} function has become an integer function,
      as it now distinguishes \emph{three} possible scenarios: no valid 
      entries, only real entries, and complex entries.
\item The \texttt{SLHAWrite} subroutine no longer has arguments for
      program name and version.  Such informational messages can now be
      added with the much more general subroutine \texttt{SLHAInfo}.
\end{itemize}


\section{Examples}
\label{sect:examples}

Consider the following example program, which just copies one SLHA file 
to another:
\begin{verbatim}
        program copy_slha_file
        implicit none

#include "SLHA.h"

        integer error
        double complex slhadata(nslhadata)

        call SLHARead(error, slhadata, "infile.slha", 0)
        if( error .ne. 0 ) stop "Read error"

        call SLHAInfo(BlockSPInfo(1), 1, "My Test Program")
        call SLHAInfo(BlockSPInfo(1), 2, "1.0")

        call SLHAWrite(error, slhadata, "outfile.slha")
        if( error .ne. 0 ) stop "Write error"
        end
\end{verbatim}
Already in this simple program a couple of things can be seen:
\begin{itemize}
\item the file \texttt{SLHA.h} must be included in every function or 
      subroutine that uses the SLHA routines and this must be done
      using the preprocessor \texttt{\#include} (not Fortran's
      \texttt{include}), thus the program file should have the
      extension \texttt{.F} (capital F).
\item \texttt{slhadata} must be declared as a double complex array of 
      length \texttt{nslhadata}.
\item One should not continue with processing if a non-zero error
      flag is returned.
\end{itemize}
A more sensible application would add something to the \texttt{slhadata} 
before writing them out again.  The next little program pretends to 
compute the fermionic Z decays (by calling a hypothetical subroutine 
\texttt{MyCalculation}) and adds them to \texttt{slhadata}:
\begin{verbatim}
        program compute_decays
        implicit none

#include "SLHA.h"
#include "PDG.h"

        integer error, decay, t, g
        double complex slhadata(nslhadata)
        double precision total_width, br(4,3)
        integer ferm_id(4,3)
        data ferm_id /
     &    PDG_nu_e, PDG_electron, PDG_up, PDG_down,
     &    PDG_nu_mu, PDG_muon, PDG_charm, PDG_strange, 
     &    PDG_nu_tau, PDG_tau, PDG_top, PDG_bottom /

        call SLHARead(error, slhadata, "infile.slha", 0)
        if( error .ne. 0 ) stop "Read error"

* compute the decays with parameters taken from the slhadata:
        call MyCalculation(SMInputs_MZ, MinPar_TB, ...,
     &    total_width, br)

        decay = SLHANewDecay(slhadata, total_width, PDG_Z)
        do g = 1, 3
          do t = 1, 4
            call SLHAAddDecay(slhadata, br(t,g), decay,
     &        2, ferm_id(t,g), -ferm_id(t,g))
          enddo
        enddo

        call SLHAInfo(BlockDCInfo(1), 1, "My Decay Calculator")
        call SLHAInfo(BlockDCInfo(1), 2, "3.1415")

        call SLHAWrite(error, slhadata, "outfile.slha")
        if( error .ne. 0 ) stop "Write error"
        end
\end{verbatim}
Demonstrated here is the access of SLHA data (\texttt{SMInputs_MZ}, 
\texttt{MinPar_TB}) and the setting of decay information.


\section{Building and Compiling}
\label{sect:build}

The SLHA library package can be downloaded as a gzipped tar archive from 
the Web site \texttt{http://www.feynarts.de/slha}.  After unpacking the 
archive, change into the directory \texttt{SLHALib-2.0} and type
\begin{verbatim}
  ./configure
  make
\end{verbatim}
Some simple demonstration programs (in the \texttt{demo} subdirectory)
are built together with the library \texttt{libSLHA.a}.

Compiling a program that uses the SLHA library is in principle equally
straightforward.  The only tricky thing is that one has to relax
Fortran's 72-column limit.  This is because even lines perfectly within
the 72-column range may become longer after the preprocessor's 
substitutions.  While essentially every Fortran compiler offers such an 
option, the name is quite different.  A glance at the man page should 
suffice to find out.  Here are a few common choices:
\begin{center}
\begin{tabular}{l|l|l}
Compiler & Platform/OS & Option name \\ \hline
g77 & any & \texttt{-ffixed-line-length-none} \\
pgf77 & Linux x86 & \texttt{-Mextend} \\
ifort & Linux x86 & \texttt{-extend_source} \\
f77 & Tru64 Alpha & \texttt{-extend_source} \\
f77 & SunOS, Solaris & \texttt{-e} \\
fort77 & HP-UX & \texttt{+es}
\end{tabular}
\end{center}
To compile and link your program, add this option and
\texttt{-I}\textit{path} \texttt{-L}\textit{path} \texttt{-lSLHA} to the
compiler command line, where \textit{path} is the location of the SLHA
library, \eg
\begin{verbatim}
pgf77 -Mextend -I$HOME/SLHALib-2.0 myprogram.F -L$HOME/SLHALib-2.0 -lSLHA
\end{verbatim}

All externally visible symbols of the SLHA library start with the prefix 
\texttt{SLHA} and should thus pretty much avoid symbol conflicts.


\section{Summary}
\label{sect:summary}

The SLHA library presented here provides simple functions to read and
write files in SLHA format.  Data are kept in a single double complex
array and accessed through preprocessor variables.  The library is
written in native Fortran 77 and is easy to build.  The source code is
openly available at \texttt{http://www.feynarts.de/slha} and is
distributed under the GNU Library General Public License.

The author welcomes any kind of feedback, in particular bug and 
performance reports, at hahn@feynarts.de.


\section*{Acknowledgements}

I thank J.~Guasch for providing a first implementation of some SLHA2
features and introducing useful routines.


\newcommand{\volyearpage}[3]{\textbf{#1} (#2) #3}
\newcommand{\cpc}{\textsl{Comp.\ Phys.\ Commun.} \volyearpage}


\begin{thebibliography}{99}

\bibitem{slha1}
P.~Skands et al., hep-ph/0311123.

\bibitem{slha2}
B.C.~Allanach et al., hep-ph/0602198.

\bibitem{slhalib1}
T. Hahn, hep-ph/0408283.

\bibitem{famssm}
T.~Hahn and C.~Schappacher, \cpc{143}{2002}{54} [hep-ph/0105349].

\end{thebibliography}
\end{document}